\documentclass[%
 aip,
 amsmath,amssymb,
 reprint,%
]{revtex4-1}

\usepackage{dcolumn}
\usepackage{bm}
\usepackage{hyperref}
\usepackage[usenames,dvipsnames]{xcolor}
\usepackage[normalem]{ulem}
\usepackage[utf8]{inputenc}
\usepackage{amsmath}
\usepackage{amsthm}
\usepackage{amsmath}
\usepackage{amsfonts}
\usepackage{blindtext}
\usepackage[utf8]{inputenc}
\usepackage[T1]{fontenc}
\usepackage{graphicx}
\usepackage{float}
\usepackage{wrapfig}
\usepackage{bm}
\usepackage{braket}
\usepackage{grffile}
\usepackage[normalem]{ulem}
\usepackage[usenames,dvipsnames]{xcolor}
\usepackage{mathptmx}

\def\RMV#1{{}}

\newcommand{\bu}{\bm{u}}
\newcommand{\bk}{\bm{k}}
\newcommand{\bx}{\bm{x}}
\newcommand{\tu}{\tilde{u}}

\newcommand{\tp}{\tilde{p}}

\begin{document}

\title{Synchronizing subgrid scale models of turbulence to data}

\author{M. Buzzicotti}
\affiliation{Dept. Physics and INFN, University of Rome ``Tor Vergata'', Rome 00133, Italy.}
\author{P. Clark Di Leoni}
\email{pato@jhu.edu}
\affiliation{Department of Mechanical Engineering, Johns Hopkins
University, Baltimore, Maryland 21218, USA.}

\date{\today}

\begin{abstract}
Large Eddy Simulations of turbulent flows are powerful tools used in
many engineering and geophysical settings. Choosing the right value of the free parameters for their subgrid scale models is a crucial task for which the current methods present several shortcomings. Using a technique called nudging we show that Large Eddy Simulations can synchronize to data coming from a high-resolution direct numerical simulation of homogeneous and isotropic turbulence. Furthermore, we found that the degree of synchronization is dependent on the value of the parameters of the subgrid scale models utilized, suggesting that nudging can be used as a way to select the best parameters for a model. For example, we show that for the Smagorinsky model synchronization is optimal when its constant takes the usual value of $0.16$. Analyzing synchronization dynamics puts the focus on reconstructing trajectories in phase space, contrary to  traditional a posteriori tests of Large Eddy Simulations where the statistics of the flows are compared.  These results open up the possibility of utilizing non-statistical analysis in a posteriori tests of Large Eddy Simulations.
\end {abstract}
\maketitle

\section{Introduction} 

Fluid turbulence, with its nonlinear and multiscale nature \cite{Davidson,Frisch1995}, is a notoriously complicated and costly problem to solve numerically in a direct way. For these reasons, a whole array of turbulence models are regularly employed for both scientific research and industrial day-to-day tasks.  Large Eddy Simulations (LES) are one particular family of models that achieve a good compromise between cost and accuracy and are widely used in engineering and geophysical applications. Their core approach consists of solving the larger scales of a fluid flow directly, while modeling the effects of the smaller eddies not resolved explicitly \cite{Meneveau00,Lesieur2005}. The term modeled is the one associated with the stresses of the unresolved (or subgrid, as they are smaller than the computational grid) motions and the models, or closures, always introduce free parameters that must be set. In some cases, theoretical arguments can be used to determine the value of these parameters, as in the Lilly-Smagorinsky model for homogeneous and isotropic turbulence \cite{Smagorinsky63,Lilly67,Pope}. The usual approach though consists of comparing the models with data coming form either experiments or Direct Numerical Simulations (DNS) where all scales are resolved.

Tests of LES can be classified into a priori~\cite{Piomelli88,buzzicotti2018} and a posteriori~\cite{Cerutti98,linkmann2018,biferale2019self}. The former approach consists of applying the models to filtered data generated in DNS or experiments to obtain a representation of how the stresses would look like in an LES and then compare against the true subgrid stresses, either by calculating the correlations between the two fields or comparing their one- \cite{Meneveau94} or two-points statistics \cite{Clark20b}.
In the latter approach, actual LES are run and the results are compared against those of DNS or experiments, but as turbulence is chaotic these comparisons are usually solely statistical, as the different phase space trajectories do not match.
Several tools coming from the data assimilation and optimization worlds are employed to effectively perform these comparisons and pick the results, such as Bayesian inference \cite{Kennedy01, Chertkov10, heas_power_2012, pang_discovering_2017, bocquet_bayesian_2020}, variational methods \cite{Rawlins07, wang_discrete_2019} and ensemble methods \cite{Anderson01, Ruiz13, Mons19}.
But given the limitations of the testing methods and the high-dimensional and chaotic nature of turbulence, the questions of what is the best choice of parameters that best models a flow and how to find that value is not always clearly answered.

We approach the issue from a synchronization perspective \cite{Carrassi08, Lalescu13} by using nudging~\cite{hoke1976init,lakshmivarahan2013nudging,clark2020synch}. Nudging is a data assimilation method \cite{Kalnay,Bauer15} that consists of adding a Newton relaxation term to the equations of motion with the purpose of nudging the flow to a prescribed field. The prescribed field is usually generated from data, either computational, experimental, or observational and the nudging term only acts where the data is available, i.e., the probe locations from where the data was extracted or the Fourier scales at which it was obtained. Nudging has been used to match boundary conditions in Numerical Weather Prediction \cite{Vonstorch00,Waldron96,Miguez-macho04} and has been shown to be capable of synchronizing a wide variety of flows to data \cite{Biswas17, Foias16, Albanez16, Pazo16, Farhat19}. Most importantly, nudging has proven to be effective at synchronizing three dimensional and fully developed turbulence \cite{clark2020synch} and shows sensitivity to the value of the physical parameters of the flow \cite{clark2019}, meaning it can detect discrepancies between the data and the nudged flow.

In this work we harness the power of nudging to show that a LES can synchronize to DNS data 
and that the level of synchronization is dependant on the choice of parameters of the models.
The flow of choice is homogeneous and isotropic turbulence and we test three different models: the classical Smagorinsky model, an Entropic Lattice Boltzmann inspired model and a Smagorinsky plus non-linear gradients model. In all three cases there exists an optimal choice of parameters that maximizes the degree of synchronization.
The paper is structured as follow, in section~\ref{sect:methods} we introduce the three subgrid scales models considered in the work, how the nudging term can be added to the LES equations and the numerical set-up used in our simulations. In section~\ref{sect:results} we present numerical results and in section~\ref{sect:conclusions} we discuss our main conclusions.

\section{Equations and methods} 
\label{sect:methods}

\subsection{Large Eddy Simulations} 

Incompressible fluid flow motion is described by the Navier-Stokes equations (NSE),

\begin{equation}
    \frac{\partial u_i}{\partial t} + u_k \frac{\partial u_i}{\partial x_k} =
    - \frac{\partial p}{\partial x_i} + \nu \frac{\partial^2
    u_i}{\partial x^2_k},
    \label{nse_idx}
\end{equation}
where $\bm{u}$ is the three dimensional velocity field, $p$ is the pressure (divided by the density) and $\nu$ is the kinematic viscosity, plus the incompressibility condition

\begin{equation}
    \frac{\partial u_i}{\partial x_i} = 0 ,
\end{equation}
and corresponding initial and boundary conditions.

Given a filtering operator $\mathcal{F}$ which filters out motions
smaller than a prescribed length $\Delta$, the LES equations are obtained by applying $\mathcal{F}$ to the NSE, resulting in

\begin{equation}
    \frac{\partial \tu_i}{\partial t} + \tu_k \frac{\partial \tu_i}{\partial x_k} =
    - \frac{\partial \tp}{\partial x_i} + \nu \frac{\partial^2
    \tu_i}{\partial x^2_k} - \frac{\partial \tau_{ik}}{\partial x_k},
    \label{nse_filt_idx}
\end{equation}
where $\tu_i =  \mathcal{F}(u_i)$, $\tp =  \mathcal{F}(p)$ and
\begin{equation}
\begin{aligned}
    \tau_{ij} = &\mathcal{F}(u_i u_j) - \mathcal{F}(u_i) \mathcal{F}(u_j)
    \\
    &- \frac13 \left( \mathcal{F}(u_i u_i) - \mathcal{F}(u_i)^2 \right) \delta_{ij},
    \label{tau_exact}
\end{aligned}
\end{equation}

\begin{equation}
    \tilde{p}=\mathcal{F}(p)  
    + \frac13 \left( \mathcal{F}(u_i u_i) - \mathcal{F}(u_i)^2 \right),
\end{equation}
are the deviatoric part of the subgrid-scale stresses and the modified pressure, respectively. The incompressibility condition simply becomes the divergence free condition but for the filtered velocity fields.

As the subgrid-scale stresses depend on the product of unfiltered
fields, the equations are closed by modeling $\tau_{ij}$ in terms of the filtered variables. In this work we analyze three different models. The first one is the classic Smagorinsky model
\begin{equation}
    \tau^{S}_{ij} = -2\nu_S \tilde{S}_{ij},
    \label{tau_smag}
\end{equation}

where
\begin{equation}
    \tilde{S}_{ij} = \frac12 \left( \frac{\partial \tu_i}{\partial x_j} + \frac{\partial
    \tu_j}{\partial x_i} \right),
\end{equation}
are the filtered strain rates and
\begin{equation}
    \nu_S = (c_S \Delta)^2 \sqrt{2 \tilde{S}_{kl} \tilde{S}_{kl}}
\end{equation}
is the eddy viscosity~\cite{Lilly67}. The model has only one free parameter, the Smagorinsky constant $c_S$. In homogeneous and isotropic turbulence this constant usually takes values around $0.16$~\cite{Meneveau00}.

The second model that we consider comes from the macroscopic formulation of the Entropic Lattice Boltzmann Model (ELBM) \cite{Karlin1999,ansumali2002single}. In particular, we define the modeled subgrid stress tensor as,
\begin{equation}
    \tau^{E}_{ij} = -2\nu_E \tilde{S}_{ij},
    \label{tau_smag}
\end{equation}
where the eddy viscosity now takes the interesting formulation derived in \cite{Malaspinas2008},

\begin{equation}
 \nu^E_e = (c_E \Delta)^2  \frac{S_{\lambda\mu}S_{\mu\gamma}S_{\gamma\lambda}}{S_{\gamma\delta}S_{\gamma\delta}},
\label{eq:nu_e}
\end{equation}
we will refer to this model as the `Entropic' model hereafter. What makes $\nu^E_e$ particularly attractive is that it is not only proportional to the stress tensor as the Smagorinsky eddy viscosity but it results also in a non-positive definite subgrid scales closure. This means that the modeled subgrid tensor $\tau^E_{ij}$ as the exact one $\tau_{ij}$ can produce backscatter events of energy going to the resolved scales and not only from the resolved to the subgrid scales with a dissipative effect~\cite{waleffe1992,fang2012time,biferale2012inverse,chen2003joint}. This model too only has one free parameter, $c_E$.

The third and last model that we considered consists in a combination between the Smagorinsky closure and the non-linear gradient or tensor eddy viscosity model, as discussed in \cite{Meneveau00},

\begin{equation}
 (c_{nl} \Delta)^2 \frac{\partial \tilde{u}_i}{\partial x_k}\frac{\partial \tilde{u}_j}{\partial x_k}.
\label{eq:nu_nl}
\end{equation}
As in the Entropic case, the gradient term is able to produce backscattering of energy.
If this term is implemented just by itself simulations do not dissipate enough energy and typically produce inaccurate results. Therefore, the purely dissipative Smagorinsky terms is added to ensure stability. Following this idea, the mixed model can be written as

\begin{equation}
 \tau^{nl}_{ij} = (c_{nl} \Delta)^2 \frac{\partial \tilde{u}_i}{\partial x_k}\frac{\partial \tilde{u}_j}{\partial x_k} - (c_S \Delta)^2 \sqrt{2 \tilde{S}_{kl} \tilde{S}_{kl}} \,\, \tilde{S}_{ij}.
\label{eq:nu_nl}
\end{equation}
Contraty to other two cases, this model has two free parameters, $c_{nl}$ and $c_S$.

Traditionally, methods to estimate the values of the free parameters of every model can be classified into a priori and a posteriori approaches. In a priori methods data coming from either a DNS or experiments is first used to calculate the exact subgrid-scale stresses $\tau_{ij}$ as defined in \eqref{tau_exact}, and then used to calculate the stresses according to the model being studied, finally the two fields are compared. While point-wise comparisons such as correlation coefficients and mean errors are commonly performed, these do not carry much prognosis power when it comes to predicting the statistic of a LES. Statistical comparisons, like comparing mean energy dissipation \cite{Meneveau94} or two-point correlations \cite{Clark20b}, have firmer foundations but still deal with filtered exact fields and not actual LES solutions. On the other hand, a posteriori approaches work by comparing the resulting flow statistics obtained by both DNS (and/or experiments) and LES. As turbulence is chaotic, the phase space trajectories coming from the different simulations can only be compared in a statistical sense. 
Methods such as ensemble Kalman filters \cite{Evensen, Houtekamer16} and variational optimization \cite{Talagrand87, Rawlins07} are then used to analyze the a posteriori simulations and find the optimal parameters.

\subsection{Nudging the LES equations} 

Nudging is a data assimilation method based on adding a penalization
term to the equations of motion with the purpose of keeping the
evolution close to some data provided. In the context of LES, the nudging protocol takes the form:

\begin{equation}
    \frac{\partial \tu_i}{\partial t} + \tu_k \frac{\partial \tu_i}{\partial x_k} =
    - \frac{\partial \tp}{\partial x_i} + \nu \frac{\partial^2
    \tu_i}{\partial x^2_k} - \frac{\partial \tau_{ki}}{\partial x_k}
    - \alpha \mathcal{I} (\tu_i - \tu^{\rm ref}_i),
    \label{nudged_eqs}
\end{equation}
where $\tilde{\bm{u}}^{\rm ref}$ is the reference data provided,
$\mathcal{I}$ is a filtering operator that acts only where data is
provided, and $\alpha$ is the magnitude of the nudging term.

Nudging will make the flow try to follow the data along the specific dynamical trajectory of the reference simulation. Nudging has been shown to be able to synchronize turbulence, both partially and fully, and is sensitive to the choice of parameters~\cite{clark2019,clark2020synch}. 

\subsection{Numerical set-up} 

The reference data was generated by solving Eq.~\eqref{nse_idx} on a periodic cubic box using a pseudospectral approach with an Adams-Bashfort two-step method for the time integration and the $2/3$ rule for dealising. The grid had $N_{data}^3=2048^3$ points, the time step was equal to $dt = 5 \cdot 10^{-5}$. The size of the computational box is $2\pi$ and the the integral length is $L=1.1$. An isotropic, constant in time forcing term, $F$, acting only on the large scales $k_f = [1, 2]$, was added to the RHS of Eq.~\eqref{nse_idx}. The characteristic speed of the flow is $U=0.92$. The viscosity used was equal to $\nu=1.6 \cdot 10^{-4}$. The resulting Reynolds number $Re = U L /\nu$ is of the order of $6300$ and eddy turnover time is $T=L/U=1.22$. 

The data from the fully resolved simulation were then filtered and used as reference, $\tu^{\rm ref}_i$, to nudge the LES equations~\eqref{nudged_eqs} with the different models. The filter operators $\mathcal{F}$ and $\mathcal{I}$ used were both identical sharp Fourier cut-off filters 

\begin{equation}
\mathcal{F}(\bu(\bx,t)) = \mathcal{I}(\bu(\bx,t)) = \sum_{|\bk| \leq k_n} \hat{\bu}(\bk,t) e^{i \bk \cdot \bx}
\label{eq:filter}
\end{equation}
with $k_n = 85$. The LES are performed with the same pseudo-spectral fully dealiased approach on the same periodic domain of size $2\pi$ but with a spatial resolution of $N^3=256^3$ grid points, resulting in a maximum wavenumber equal to $k_n$. 
Nudging does not need to have reference information available at every scale in order to work in more general settings \cite{clark2020synch}, but as we are interested in analyzing how the different scales synchronize and as the subgrid models activity is concentrated in the smaller scales, we nudge over the whole range of resolved scales.
All simulations are forced with the same isotropic constant forcing mechanisms acting only on the large system scales ($k_f = [1, 2]$). The choice of using the same deterministic forcing mechanism in both LES and reference simulations allows us to highlight the effects produced by the modeling of the subgrid term on the dynamics of the LES resolved scales. The value of the nudging amplitude $\alpha$ was always set to $10$ except when noted. The first snapshot of the fully resolved simulation, filtered as described in Eq.~\ref{eq:filter}, was used as an initial condition in all LES simulations.

\section{Results} 
\label{sect:results}

\subsection{Smagorinsky model}

\begin{figure}%
    \includegraphics[width=0.45\textwidth]{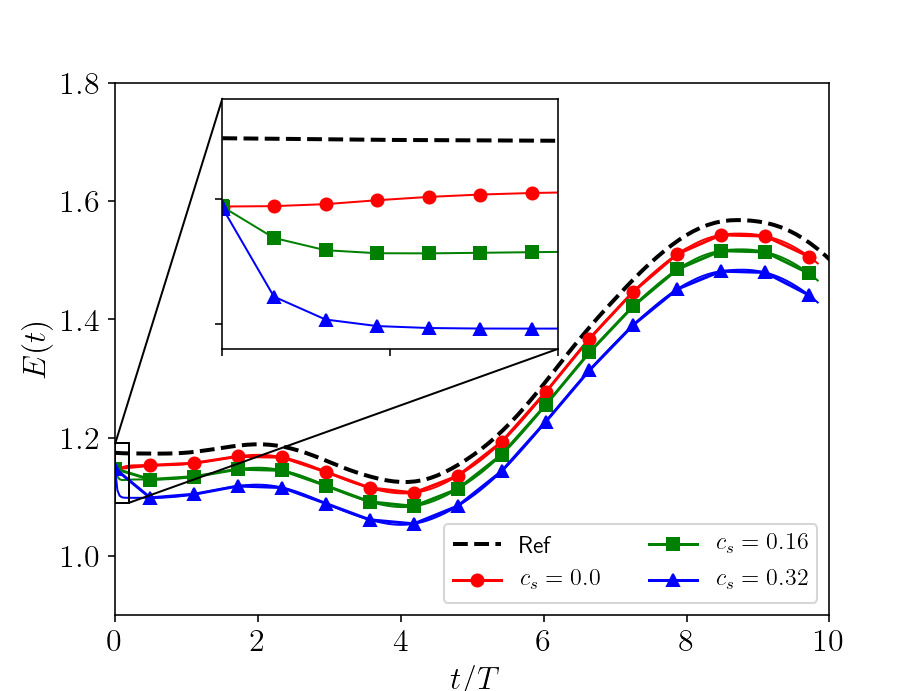}
    \caption{Energy evolution in time for the reference simulation at a resolution of $2048^3$ grid points (black dashed line) and for three different LES simulations equipped with the Smagorinsky closure at changing of the model free parameter, $c_S$.}
    \label{fig:ener_time}
\end{figure}

In Fig.~\ref{fig:ener_time} we show the evolution of the kinetic energy of the reference simulation and of three nudged Smagorinsky simulations with different values of $c_S$. The figure inset shows a close up to the very early stages of the simulations. As the nudging acts on every resolved scale, none of the three nudged simulations have any trouble synchronizing, i.e. following, the reference simulation, although they all have distinct and different evolutions. All nudged simulations have a total energy which is slightly lower than the reference case, as was already observed in~\cite{clark2020synch}.  Let us stress that the nudging term allows the LES simulation to follow the reference data also when there is no subgrid scales model, see the red curve in Fig.~\ref{fig:ener_time} for $c_S=0$, acting itself as closure for the subgrid dynamics. The question then is: for which value of the free parameter the LES model synchronizes better?

\begin{figure*}%
    \includegraphics[width=0.45\textwidth]{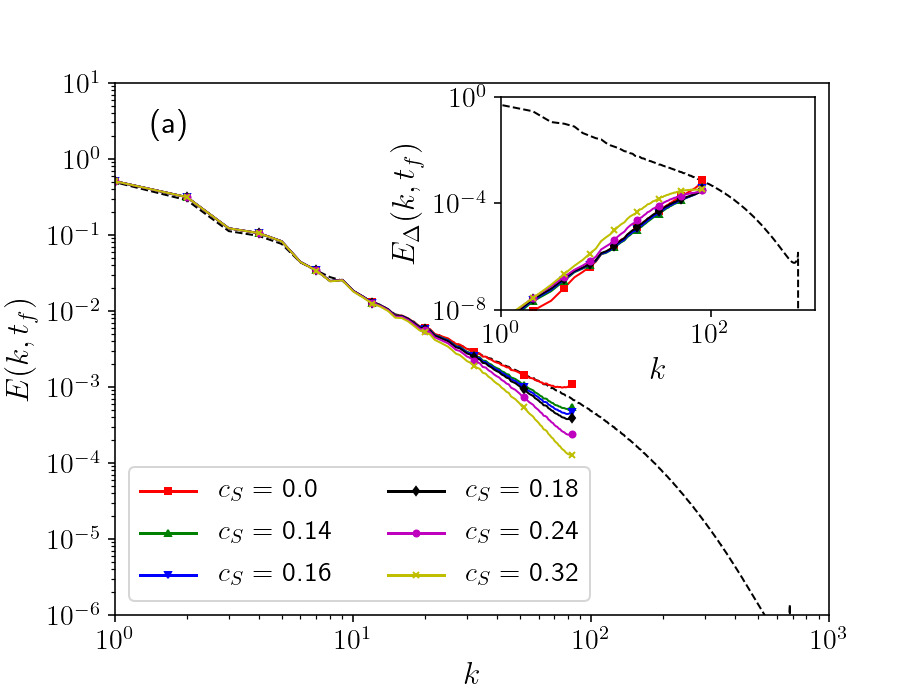}   
    \includegraphics[width=0.45\textwidth]{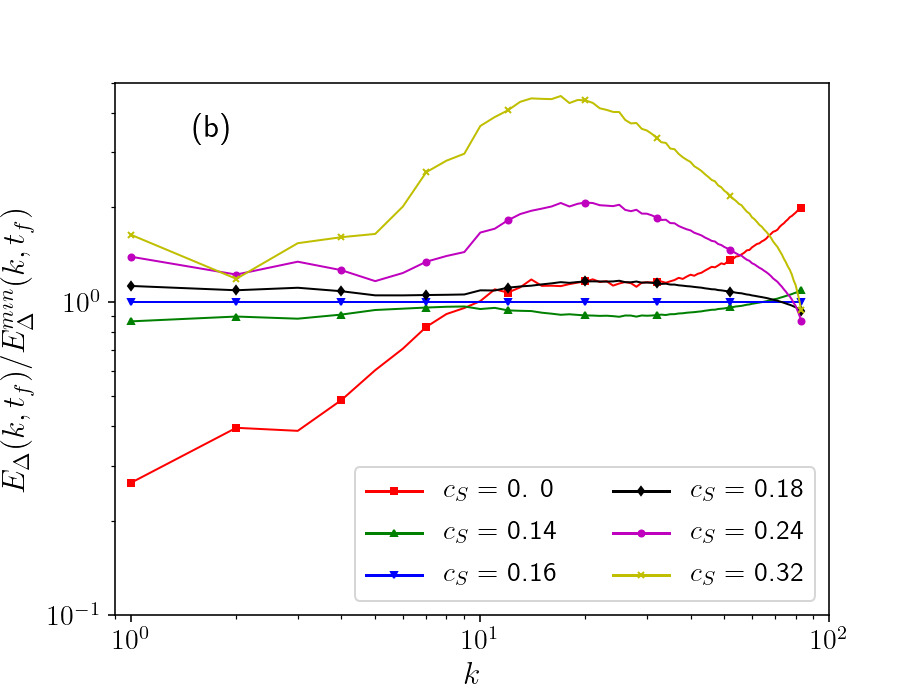}   
    \caption{Panel (a): Energy spectra, $E(k,t_f)$, of the reference simulation at a resolution of $2048^3$ grid points (black dashed line) and for the nudged LES  simulations with a Smagorinsky closure at a resolution of $256^3$ grid points for different values of $c_S$. In the inset of the same figure we show the spectra of the difference, $E_\Delta(k, t_f)$, measured for the same simulations of the main panel and compared with the energy spectra of the reference simulation.
    Panel (b): The spectra of the difference normalized by the $E_\Delta(k, t_f)$ measured at $c_S=0.16$, the blue lines with downward triangles is equal to $1$ by definition.}
    \label{fig:spectra_smag}
\end{figure*}
\begin{figure}%
    \includegraphics[width=0.45\textwidth]{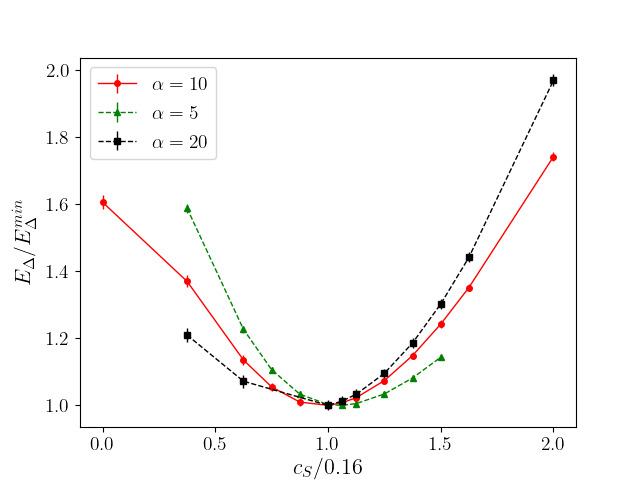}   
    \caption{Time averaged relative errors, $E_\Delta$, for different amplitudes of the free parameter in the Smagorinsky model, $c_S$. The x- and y-axis are normalized respectively by $0.16$ and the relative error measured at $c_S=0.16$. The LES simulations where performed with different amplitudes of the nudging term, $\alpha=5$ (green line with triangles), $\alpha=10$ (red line with circles) and $\alpha=20$ (black line with squares).}
    \label{fig:scan_coef_smag}
\end{figure}

In Fig.~\ref{fig:spectra_smag}(a) we show the energy spectra,

\begin{equation}
    E (k, t) = \frac12 \sum_{k\leq |\bk| \leq k+1} \vert \hat{\bu}(\bk, t) \vert^2,
\end{equation}
at the final time of the simulations, $t_f$ (see Fig.~\ref{fig:ener_time}) for the reference flow and six nudged Smagorinsky simulations with different values of $c_S$. The spectra of the different nudged cases all coincide with the reference over almost the whole range of scales, differences between the different runs can be appreciated only at the highest wavenumbers. While the nudged simulation with $c_S=0$ is the one whose spectra is the closest to the reference one, it is important to remember that here we are only comparing the total amount of energy at each scale, not the actual configuration. To that end we calculate the spectra of the differences

\begin{figure*}%
    \includegraphics[width=0.45\textwidth]{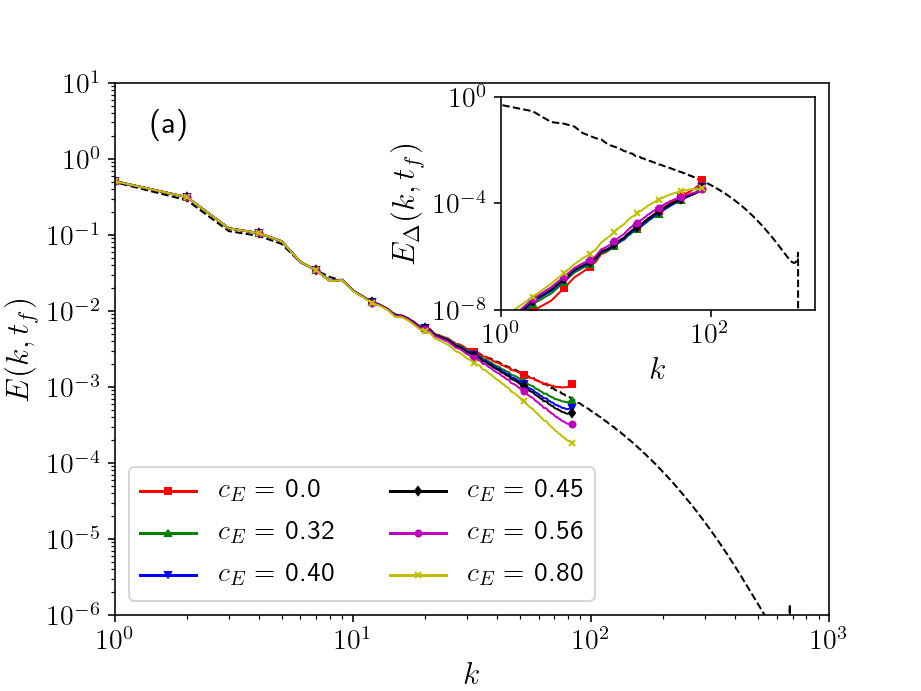}   
    \includegraphics[width=0.45\textwidth]{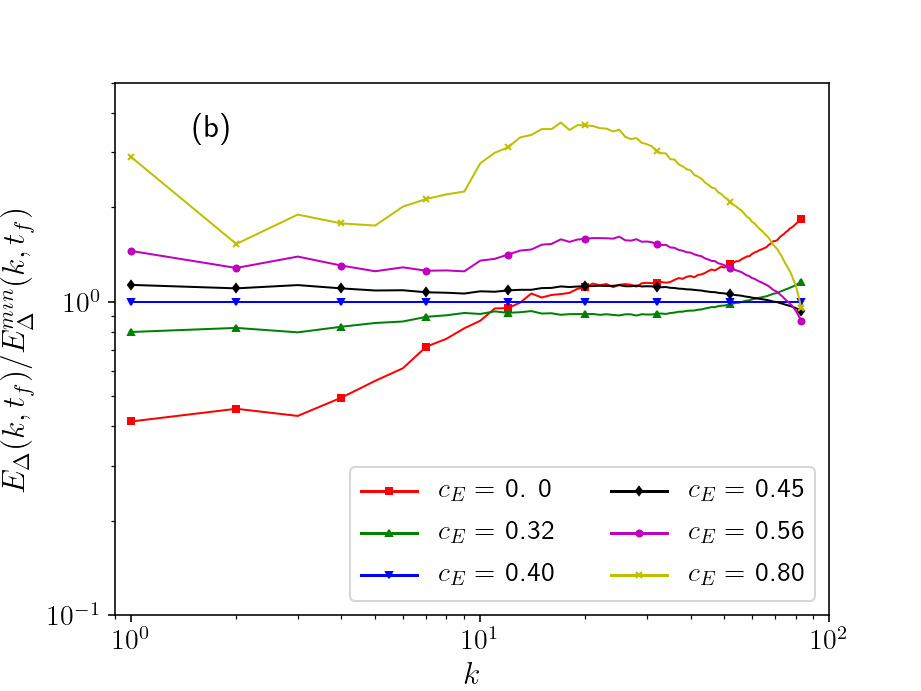}   
    \caption{Panel (a): Energy spectra, $E(k,t_f)$, of the reference simulation at a resolution of $2048^3$ grid points (black dashed line) and for the nudged LES  simulations with the Entropic model closure at a resolution of $256^3$ grid points for different values of $c_E$. In the inset of the same figure we show the spectra of the difference, $E_\Delta(k, t_f)$, measured for the same simulations of the main panel and compared with the energy spectra of the reference simulation.
    Panel (b): The spectra of the difference normalized by the $E_\Delta(k, t_f)$ measured at $c_E=0.40$, the blue lines with downward triangles is equal to $1$ by definition.}
    \label{fig:spectra_malas}
\end{figure*}

\begin{figure}%
    \includegraphics[width=0.45\textwidth]{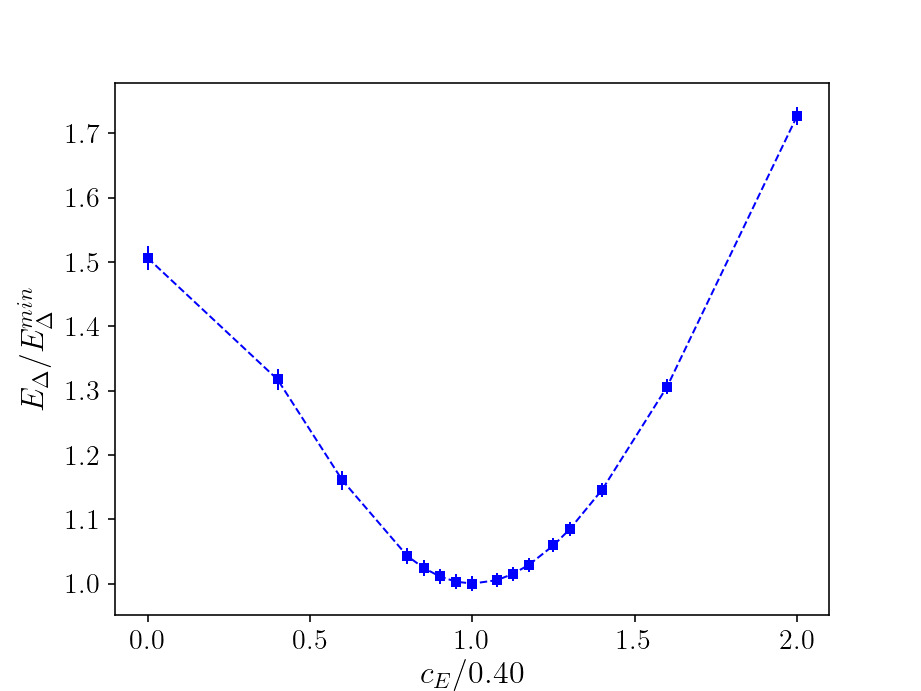}   
    \caption{Time averaged relative errors, $E_\Delta$, for different amplitudes of the free parameter in the Entropic model, $c_E$. The x- and y-axis are normalized respectively by $0.40$ and the relative error measured at $c_E=0.40$. The LES simulations where performed with nudging amplitude $\alpha=10$.}
    \label{fig:scan_coeff_malas}
\end{figure}

\begin{equation}
    E_\Delta (k, t) = \frac12 \sum_{k\leq |\bk| \leq k+1} \vert \hat{\bu}(\bk,t) - \hat{\bu}^{\rm ref}(\bk,t) \vert^2,
\end{equation}
and show it in Fig.~\ref{fig:spectra_smag}(a) alongside the spectra of the reference simulation (evaluted at final time $t_f$), while in Fig.~\ref{fig:spectra_smag}(b) we show the same spectra (except for the reference one) but all normalized by the spectrum of the case with $c_S=0.16$, whose integral, i.e., total error, was found to be the smallest, as we discuss in detail below. Notice that the quantity $E_\Delta (k, t)$ tracks the kinetic energy of the difference of the two fields, not the difference of the energies of the two fields, and, as such, also takes into consideration any phase difference the Fourier coefficients of the two fields may have. These two figures show us how the models with lower values of $c_S$ do a slightly better job at synchronizing the largest scales of the flow but have problems at the smallest scales (compared to the case with $c_S=0.16$). On the other hand, simulations with a very high value of $c_S$ can have problems at all scales. This result suggests that the larger scales are less affected by the absence of the subgrid scales in the LES simulations, and their dynamics do not require any model. On the other hand, moving towards the filter cutoff, the missing degrees of freedom devolve on a larger deviation in the dynamics of such scales and the model becomes necessary. Let us stress once more that having found that the modelled LES can be synchronized to fully resolved simulations, now our aim to find the optimal parameters choice that maximizes the level of synchronization.

Finally, in Fig.~\ref{fig:scan_coef_smag} we make a global quantitative assessment of how well each simulation is synchronizing to the reference by showing the values of the time averaged relative errors $E_\Delta = \langle \sum_{1<k<k_\eta} E_\Delta(k, t) \rangle$ (where $\langle.\rangle$ denotes the time average) normalized by the value of $E_\Delta$ for the case with $c_S=0.16$ scanned over a range of $c_S$ and using different values of the nudging amplitude $\alpha$. Errorbars in the figure were obtained by calculating the standard deviation of $\sum_{1<k<k_\eta} E_\Delta(k,t)$. The dependence on $\alpha$ can be explained as follows. When $\alpha$ is too small the nudging term is less effective and the LES model is required to ensure stability at small scales, indeed, the error grows quickly when $c_S$ is too small (see green line with triangles). On the contrary, when $\alpha$ is large, the small scales stability is guaranteed by nudging also very small $c_S$, however in this case the simulations show larger deviations when increasing $c_S$, suggesting that synchronization becomes more sensitive to the error introduced at large scales by the model (see black line with squares).
For all three scans, the lowest values are encountered when $c_S \approx 0.16$, in agreement with other statistical estimations based on a priori and a posteriori tests and on theoretical arguments~\cite{Meneveau00}. As expected too, the minimum around $0.16$ is not extremely sharp, as LES are known to produce accurate results with slightly different values of $c_S$.

\subsection{Entropic model}

\begin{figure}%
    \includegraphics[width=0.45\textwidth]{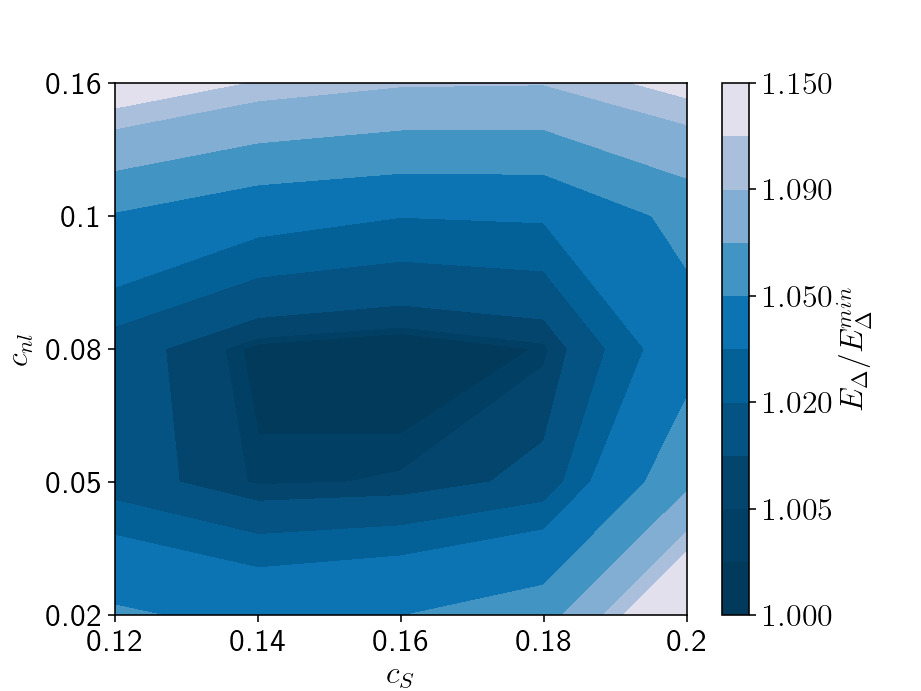}   
    \caption{Time averaged relative error, normalized to its minimum, as a function of the non-linear model parameters, $(c_S, c_{nl})$.}
    \label{fig:scan_coef_nl}
\end{figure}

We now repeat the same analysis as above but for the Entropic model outlined in  Eq.~\eqref{eq:nu_e}. In Fig.~\ref{fig:spectra_malas}(a) we show the energy spectra at the final time of the simulations for the reference flow and six nudged Entropic simulations with different values of $c_E$. Again, the spectra of the different nudged cases all coincide with the reference over almost the whole range of scales, differences between the different runs can be appreciated only at the highest wavenumbers. In the inset of the same figure we show the spectra of the differences $E_\Delta(k, t_f)$, alongside the spectra of the reference simulation, while in Fig.~\ref{fig:spectra_malas}(b) we show the same spectra (except for the reference one) but normalized by the spectrum of the case with $c_E=0.40$. The results are very similar to what is observed for the Smagorinsky model, indicating that synchronization between LES and fully resolved data is possible not only for the specific case of the Smagorinsky model.

In Fig.~\ref{fig:scan_coeff_malas} we show a scan of the time averaged total error $E_\Delta$ as a function of $c_E$. As with the previous case, the x- and y-axis are normalized respectively by the value of $c_E=0.40$ and its corresponding error, that is the value able to synchronize more closely to the reference data. Let us stress that while in the Smagorinsky case nudging provided us a result in agreement with the previous literature, this is the first time that the Entropic model has been used in a LES simulation, proving how nudging can be used to fit free parameters in new models.

\subsection{Smagorinsky model with nonlinear gradients}

Before concluding, we analyze the nudged LES simulations performed using the non-linear model, namely the Smagorinsky model plus the gradients term, outlined in Eq.~\eqref{eq:nu_nl}. The reasons for choosing this third model are twofold. First, we wanted to investigate the possibility for nudged LES to synchronize with reference data also in the presence of more complex/non-linear models. Second, the non-linear model has two free parameters and it is interesting to understand if synchronization shows a minimum also in this 2-dimensional phase-space, $(c_S, c_{nl})$. Results are summarized in Fig.~\ref{fig:scan_coef_nl}, where we show the time averaged relative errors, $E_\Delta$, measured scanning the $(c_S, c_{nl})$-space. We can conclude that also in this case the LES were able to synchronize with respect to the reference data, and that the relative averaged error presents a clear minimum around the point $(c_S=0.16, c_{nl}=0.07)$. This result further support the observation that LES can synchronize with reference data independently of the specific properties of model used.

\section{Conclusions} 
\label{sect:conclusions}

We have shown that it is possible to synchronize a LES to data coming from a fully resolved simulation of homogeneous and isotropic turbulence, meaning that the LES can follow the particular trajectory of a fully resolved case. The analysis was repeated using three different models yielding successful results in all cases. Even more interestingly, we have shown that by varying the model parameters it is possible to change the level of synchronization between the LES and data, and that for each model there exists an optimal choice of parameters that minimizes the distance between the LES and the data.  Therefore, nudging offers a way to get an estimate of the optimal model parameters that relies on reconstructing specific trajectories in phase space. We have also shown how nudging allows to have a scale-by-scale estimation of the errors introduced by the models considered. These results open up new ways in which to test LES models and choose their parameters. While in its current form the nudging algorithm presented lacks a proper optimization procedure to iterate over the parameters, the powers of synchronization could also be harnessed by different data assimilation and optimization methods.

\begin{acknowledgments}
The authors acknowledge partial funding from the
European Research Council under the European
Community’s Seventh Framework Program, ERC Grant
Agreement No. 339032.
The authors would like to thank Luca Biferale for support and inspiration. P. C. would like to thank Charles Meneveau and Tamer Zaki for useful discussions.
\end{acknowledgments}

\section*{Data Availability}
The data that support the findings of this study are available from the corresponding author upon reasonable request.

\bibliography{biblio}

\begin{thebibliography}{50}%
\makeatletter
\providecommand \@ifxundefined [1]{%
 \@ifx{#1\undefined}
}%
\providecommand \@ifnum [1]{%
 \ifnum #1\expandafter \@firstoftwo
 \else \expandafter \@secondoftwo
 \fi
}%
\providecommand \@ifx [1]{%
 \ifx #1\expandafter \@firstoftwo
 \else \expandafter \@secondoftwo
 \fi
}%
\providecommand \natexlab [1]{#1}%
\providecommand \enquote  [1]{``#1''}%
\providecommand \bibnamefont  [1]{#1}%
\providecommand \bibfnamefont [1]{#1}%
\providecommand \citenamefont [1]{#1}%
\providecommand \href@noop [0]{\@secondoftwo}%
\providecommand \href [0]{\begingroup \@sanitize@url \@href}%
\providecommand \@href[1]{\@@startlink{#1}\@@href}%
\providecommand \@@href[1]{\endgroup#1\@@endlink}%
\providecommand \@sanitize@url [0]{\catcode `\\12\catcode `\$12\catcode
  `\&12\catcode `\#12\catcode `\^12\catcode `\_12\catcode `\%12\relax}%
\providecommand \@@startlink[1]{}%
\providecommand \@@endlink[0]{}%
\providecommand \url  [0]{\begingroup\@sanitize@url \@url }%
\providecommand \@url [1]{\endgroup\@href {#1}{\urlprefix }}%
\providecommand \urlprefix  [0]{URL }%
\providecommand \Eprint [0]{\href }%
\providecommand \doibase [0]{http://dx.doi.org/}%
\providecommand \selectlanguage [0]{\@gobble}%
\providecommand \bibinfo  [0]{\@secondoftwo}%
\providecommand \bibfield  [0]{\@secondoftwo}%
\providecommand \translation [1]{[#1]}%
\providecommand \BibitemOpen [0]{}%
\providecommand \bibitemStop [0]{}%
\providecommand \bibitemNoStop [0]{.\EOS\space}%
\providecommand \EOS [0]{\spacefactor3000\relax}%
\providecommand \BibitemShut  [1]{\csname bibitem#1\endcsname}%
\let\auto@bib@innerbib\@empty
\bibitem [{\citenamefont {Davidson}(2013)}]{Davidson}%
  \BibitemOpen
  \bibfield  {author} {\bibinfo {author} {\bibfnamefont {P.~A.}\ \bibnamefont
  {Davidson}},\ }\href@noop {} {\emph {\bibinfo {title} {Turbulence in
  Rotating, Stratified and Electrically Conducting Fluids}}}\ (\bibinfo
  {publisher} {Cambridge University Press},\ \bibinfo {year}
  {2013})\BibitemShut {NoStop}%
\bibitem [{\citenamefont {Frisch}(1995)}]{Frisch1995}%
  \BibitemOpen
  \bibfield  {author} {\bibinfo {author} {\bibfnamefont {U.}~\bibnamefont
  {Frisch}},\ }\href
  {https://books.google.it/books?hl=en{\&}lr={\&}id=g-p-AwAAQBAJ{\&}oi=fnd{\&}pg=PR11{\&}dq=frisch+turbulence+legacy+citation{\&}ots=IppiBDlL43{\&}sig=fwzj4uhirqK7TbxI8X--wv9M1TY{\#}v=onepage{\&}q=frisch
  turbulence legacy citation{\&}f=false} {\emph {\bibinfo {title} {{Turbulence
  : the legacy of A.N. Kolmogorov}}}}\ (\bibinfo  {publisher} {Cambridge
  University Press},\ \bibinfo {year} {1995})\ p.\ \bibinfo {pages}
  {296}\BibitemShut {NoStop}%
\bibitem [{\citenamefont {Meneveau}\ and\ \citenamefont
  {Katz}(2000)}]{Meneveau00}%
  \BibitemOpen
  \bibfield  {author} {\bibinfo {author} {\bibfnamefont {C.}~\bibnamefont
  {Meneveau}}\ and\ \bibinfo {author} {\bibfnamefont {J.}~\bibnamefont
  {Katz}},\ }\bibfield  {title} {\enquote {\bibinfo {title} {Scale-{Invariance}
  and {Turbulence} {Models} for {Large}-{Eddy} {Simulation}},}\ }\href
  {\doibase 10.1146/annurev.fluid.32.1.1} {\bibfield  {journal} {\bibinfo
  {journal} {Annual Review of Fluid Mechanics}\ }\textbf {\bibinfo {volume}
  {32}},\ \bibinfo {pages} {1--32} (\bibinfo {year} {2000})}\BibitemShut
  {NoStop}%
\bibitem [{\citenamefont {Lesieur}, \citenamefont {M{\'e}tais},\ and\
  \citenamefont {Comte}(2005)}]{Lesieur2005}%
  \BibitemOpen
  \bibfield  {author} {\bibinfo {author} {\bibfnamefont {M.}~\bibnamefont
  {Lesieur}}, \bibinfo {author} {\bibfnamefont {O.}~\bibnamefont {M{\'e}tais}},
  \ and\ \bibinfo {author} {\bibfnamefont {P.}~\bibnamefont {Comte}},\
  }\href@noop {} {\emph {\bibinfo {title} {Large-eddy simulations of
  turbulence}}}\ (\bibinfo  {publisher} {Cambridge University Press},\ \bibinfo
  {year} {2005})\BibitemShut {NoStop}%
\bibitem [{\citenamefont {Smagorinsky}(1963)}]{Smagorinsky63}%
  \BibitemOpen
  \bibfield  {author} {\bibinfo {author} {\bibfnamefont {J.}~\bibnamefont
  {Smagorinsky}},\ }\bibfield  {title} {\enquote {\bibinfo {title} {General
  circulation experiments with the primitive equations: I. the basic
  experiment.}}\ }\href@noop {} {\bibfield  {journal} {\bibinfo  {journal}
  {Monthly weather review}\ }\textbf {\bibinfo {volume} {91}},\ \bibinfo
  {pages} {99--164} (\bibinfo {year} {1963})}\BibitemShut {NoStop}%
\bibitem [{\citenamefont {Lilly}(1967)}]{Lilly67}%
  \BibitemOpen
  \bibfield  {author} {\bibinfo {author} {\bibfnamefont {D.~K.}\ \bibnamefont
  {Lilly}},\ }\bibfield  {title} {\enquote {\bibinfo {title} {The
  representation of small scale turbulence in numerical simulation
  experiments},}\ }in\ \href@noop {} {\emph {\bibinfo {booktitle} {Proc. IBM
  Scientific Computing Symposium on environmental sciences}}},\ \bibinfo
  {editor} {edited by\ \bibinfo {editor} {\bibfnamefont {H.~H.}\ \bibnamefont
  {Goldstine}}}\ (\bibinfo {year} {1967})\ pp.\ \bibinfo {pages}
  {195--210}\BibitemShut {NoStop}%
\bibitem [{\citenamefont {Pope}(2000)}]{Pope}%
  \BibitemOpen
  \bibfield  {author} {\bibinfo {author} {\bibfnamefont {S.~B.}\ \bibnamefont
  {Pope}},\ }\href@noop {} {\emph {\bibinfo {title} {Turbulent Flows}}}\
  (\bibinfo  {publisher} {Cambridge University Press},\ \bibinfo {year}
  {2000})\BibitemShut {NoStop}%
\bibitem [{\citenamefont {Piomelli}, \citenamefont {Moin},\ and\ \citenamefont
  {Ferziger}(1988)}]{Piomelli88}%
  \BibitemOpen
  \bibfield  {author} {\bibinfo {author} {\bibfnamefont {U.}~\bibnamefont
  {Piomelli}}, \bibinfo {author} {\bibfnamefont {P.}~\bibnamefont {Moin}}, \
  and\ \bibinfo {author} {\bibfnamefont {J.~H.}\ \bibnamefont {Ferziger}},\
  }\bibfield  {title} {\enquote {\bibinfo {title} {Model consistency in large
  eddy simulation of turbulent channel flows},}\ }\href@noop {} {\bibfield
  {journal} {\bibinfo  {journal} {Phys. Fluids}\ }\textbf {\bibinfo {volume}
  {31}},\ \bibinfo {pages} {1884--1891} (\bibinfo {year} {1988})}\BibitemShut
  {NoStop}%
\bibitem [{\citenamefont {Buzzicotti}\ \emph {et~al.}(2018)\citenamefont
  {Buzzicotti}, \citenamefont {Linkmann}, \citenamefont {Aluie}, \citenamefont
  {Biferale}, \citenamefont {Brasseur},\ and\ \citenamefont
  {Meneveau}}]{buzzicotti2018}%
  \BibitemOpen
  \bibfield  {author} {\bibinfo {author} {\bibfnamefont {M.}~\bibnamefont
  {Buzzicotti}}, \bibinfo {author} {\bibfnamefont {M.}~\bibnamefont
  {Linkmann}}, \bibinfo {author} {\bibfnamefont {H.}~\bibnamefont {Aluie}},
  \bibinfo {author} {\bibfnamefont {L.}~\bibnamefont {Biferale}}, \bibinfo
  {author} {\bibfnamefont {J.}~\bibnamefont {Brasseur}}, \ and\ \bibinfo
  {author} {\bibfnamefont {C.}~\bibnamefont {Meneveau}},\ }\bibfield  {title}
  {\enquote {\bibinfo {title} {Effect of filter type on the statistics of
  energy transfer between resolved and subfilter scales from a-priori analysis
  of direct numerical simulations of isotropic turbulence},}\ }\href@noop {}
  {\bibfield  {journal} {\bibinfo  {journal} {Journal of Turbulence}\ }\textbf
  {\bibinfo {volume} {19}},\ \bibinfo {pages} {167--197} (\bibinfo {year}
  {2018})}\BibitemShut {NoStop}%
\bibitem [{\citenamefont {Cerutti}\ and\ \citenamefont
  {Meneveau}(1998)}]{Cerutti98}%
  \BibitemOpen
  \bibfield  {author} {\bibinfo {author} {\bibfnamefont {S.}~\bibnamefont
  {Cerutti}}\ and\ \bibinfo {author} {\bibfnamefont {C.}~\bibnamefont
  {Meneveau}},\ }\bibfield  {title} {\enquote {\bibinfo {title} {{Intermittency
  and relative scaling of subgrid-scale energy dissipation in isotropic
  turbulence}},}\ }\href@noop {} {\bibfield  {journal} {\bibinfo  {journal}
  {Phys. Fluids}\ }\textbf {\bibinfo {volume} {10}},\ \bibinfo {pages} {928}
  (\bibinfo {year} {1998})}\BibitemShut {NoStop}%
\bibitem [{\citenamefont {Linkmann}, \citenamefont {Buzzicotti},\ and\
  \citenamefont {Biferale}(2018)}]{linkmann2018}%
  \BibitemOpen
  \bibfield  {author} {\bibinfo {author} {\bibfnamefont {M.}~\bibnamefont
  {Linkmann}}, \bibinfo {author} {\bibfnamefont {M.}~\bibnamefont
  {Buzzicotti}}, \ and\ \bibinfo {author} {\bibfnamefont {L.}~\bibnamefont
  {Biferale}},\ }\bibfield  {title} {\enquote {\bibinfo {title} {Multi-scale
  properties of large eddy simulations: correlations between resolved-scale
  velocity-field increments and subgrid-scale quantities},}\ }\href@noop {}
  {\bibfield  {journal} {\bibinfo  {journal} {Journal of Turbulence}\ }\textbf
  {\bibinfo {volume} {19}},\ \bibinfo {pages} {493--527} (\bibinfo {year}
  {2018})}\BibitemShut {NoStop}%
\bibitem [{\citenamefont {Biferale}\ \emph {et~al.}(2019)\citenamefont
  {Biferale}, \citenamefont {Bonaccorso}, \citenamefont {Buzzicotti},\ and\
  \citenamefont {Iyer}}]{biferale2019self}%
  \BibitemOpen
  \bibfield  {author} {\bibinfo {author} {\bibfnamefont {L.}~\bibnamefont
  {Biferale}}, \bibinfo {author} {\bibfnamefont {F.}~\bibnamefont
  {Bonaccorso}}, \bibinfo {author} {\bibfnamefont {M.}~\bibnamefont
  {Buzzicotti}}, \ and\ \bibinfo {author} {\bibfnamefont {K.~P.}\ \bibnamefont
  {Iyer}},\ }\bibfield  {title} {\enquote {\bibinfo {title} {Self-similar
  subgrid-scale models for inertial range turbulence and accurate measurements
  of intermittency},}\ }\href@noop {} {\bibfield  {journal} {\bibinfo
  {journal} {Physical review letters}\ }\textbf {\bibinfo {volume} {123}},\
  \bibinfo {pages} {014503} (\bibinfo {year} {2019})}\BibitemShut {NoStop}%
\bibitem [{\citenamefont {Meneveau}(1994)}]{Meneveau94}%
  \BibitemOpen
  \bibfield  {author} {\bibinfo {author} {\bibfnamefont {C.}~\bibnamefont
  {Meneveau}},\ }\bibfield  {title} {\enquote {\bibinfo {title} {{Statistics of
  turbulence subgrid-scale stresses: Necessary conditions and experimental
  tests}},}\ }\href@noop {} {\bibfield  {journal} {\bibinfo  {journal} {Phys.
  Fluids}\ }\textbf {\bibinfo {volume} {6}},\ \bibinfo {pages} {815} (\bibinfo
  {year} {1994})}\BibitemShut {NoStop}%
\bibitem [{\citenamefont {Di~Leoni}\ \emph {et~al.}(2020)\citenamefont
  {Di~Leoni}, \citenamefont {Zaki}, \citenamefont {Karniadakis},\ and\
  \citenamefont {Meneveau}}]{Clark20b}%
  \BibitemOpen
  \bibfield  {author} {\bibinfo {author} {\bibfnamefont {P.~C.}\ \bibnamefont
  {Di~Leoni}}, \bibinfo {author} {\bibfnamefont {T.~A.}\ \bibnamefont {Zaki}},
  \bibinfo {author} {\bibfnamefont {G.}~\bibnamefont {Karniadakis}}, \ and\
  \bibinfo {author} {\bibfnamefont {C.}~\bibnamefont {Meneveau}},\ }\bibfield
  {title} {\enquote {\bibinfo {title} {Two-point stress-strain rate correlation
  structure and non-local eddy viscosity in turbulent flows},}\ }\href@noop {}
  {\bibfield  {journal} {\bibinfo  {journal} {arXiv preprint arXiv:2006.02280}\
  } (\bibinfo {year} {2020})}\BibitemShut {NoStop}%
\bibitem [{\citenamefont {Kennedy}\ and\ \citenamefont
  {O'Hagan}(2001)}]{Kennedy01}%
  \BibitemOpen
  \bibfield  {author} {\bibinfo {author} {\bibfnamefont {M.~C.}\ \bibnamefont
  {Kennedy}}\ and\ \bibinfo {author} {\bibfnamefont {A.}~\bibnamefont
  {O'Hagan}},\ }\bibfield  {title} {\enquote {\bibinfo {title} {Bayesian
  calibration of computer models},}\ }\href {\doibase 10.1111/1467-9868.00294}
  {\bibfield  {journal} {\bibinfo  {journal} {Journal of the Royal Statistical
  Society: Series B (Statistical Methodology)}\ }\textbf {\bibinfo {volume}
  {63}},\ \bibinfo {pages} {425--464} (\bibinfo {year} {2001})}\BibitemShut
  {NoStop}%
\bibitem [{\citenamefont {Chertkov}\ \emph {et~al.}(2010)\citenamefont
  {Chertkov}, \citenamefont {Kroc}, \citenamefont {Krzakala}, \citenamefont
  {Vergassola},\ and\ \citenamefont {Zdeborová}}]{Chertkov10}%
  \BibitemOpen
  \bibfield  {author} {\bibinfo {author} {\bibfnamefont {M.}~\bibnamefont
  {Chertkov}}, \bibinfo {author} {\bibfnamefont {L.}~\bibnamefont {Kroc}},
  \bibinfo {author} {\bibfnamefont {F.}~\bibnamefont {Krzakala}}, \bibinfo
  {author} {\bibfnamefont {M.}~\bibnamefont {Vergassola}}, \ and\ \bibinfo
  {author} {\bibfnamefont {L.}~\bibnamefont {Zdeborová}},\ }\bibfield  {title}
  {\enquote {\bibinfo {title} {Inference in particle tracking experiments by
  passing messages between images},}\ }\href {\doibase 10.1073/pnas.0910994107}
  {\bibfield  {journal} {\bibinfo  {journal} {Proceedings of the National
  Academy of Sciences}\ }\textbf {\bibinfo {volume} {107}},\ \bibinfo {pages}
  {7663--7668} (\bibinfo {year} {2010})}\BibitemShut {NoStop}%
\bibitem [{\citenamefont {Héas}\ \emph {et~al.}(2012)\citenamefont {Héas},
  \citenamefont {Mémin}, \citenamefont {Heitz},\ and\ \citenamefont
  {Mininni}}]{heas_power_2012}%
  \BibitemOpen
  \bibfield  {author} {\bibinfo {author} {\bibfnamefont {P.}~\bibnamefont
  {Héas}}, \bibinfo {author} {\bibfnamefont {E.}~\bibnamefont {Mémin}},
  \bibinfo {author} {\bibfnamefont {D.}~\bibnamefont {Heitz}}, \ and\ \bibinfo
  {author} {\bibfnamefont {P.~D.}\ \bibnamefont {Mininni}},\ }\bibfield
  {title} {\enquote {\bibinfo {title} {Power laws and inverse motion modelling:
  application to turbulence measurements from satellite images},}\ }\href
  {\doibase 10.3402/tellusa.v64i0.10962} {\bibfield  {journal} {\bibinfo
  {journal} {Tellus A: Dynamic Meteorology and Oceanography}\ }\textbf
  {\bibinfo {volume} {64}},\ \bibinfo {pages} {10962} (\bibinfo {year}
  {2012})}\BibitemShut {NoStop}%
\bibitem [{\citenamefont {Pang}\ \emph {et~al.}(2017)\citenamefont {Pang},
  \citenamefont {Perdikaris}, \citenamefont {Cai},\ and\ \citenamefont
  {Karniadakis}}]{pang_discovering_2017}%
  \BibitemOpen
  \bibfield  {author} {\bibinfo {author} {\bibfnamefont {G.}~\bibnamefont
  {Pang}}, \bibinfo {author} {\bibfnamefont {P.}~\bibnamefont {Perdikaris}},
  \bibinfo {author} {\bibfnamefont {W.}~\bibnamefont {Cai}}, \ and\ \bibinfo
  {author} {\bibfnamefont {G.~E.}\ \bibnamefont {Karniadakis}},\ }\bibfield
  {title} {\enquote {\bibinfo {title} {Discovering variable fractional orders
  of advection–dispersion equations from field data using multi-fidelity
  {Bayesian} optimization},}\ }\href {\doibase 10.1016/j.jcp.2017.07.052}
  {\bibfield  {journal} {\bibinfo  {journal} {Journal of Computational
  Physics}\ }\textbf {\bibinfo {volume} {348}},\ \bibinfo {pages} {694--714}
  (\bibinfo {year} {2017})}\BibitemShut {NoStop}%
\bibitem [{\citenamefont {Bocquet}\ \emph {et~al.}(2020)\citenamefont
  {Bocquet}, \citenamefont {Brajard}, \citenamefont {Carrassi},\ and\
  \citenamefont {Bertino}}]{bocquet_bayesian_2020}%
  \BibitemOpen
  \bibfield  {author} {\bibinfo {author} {\bibfnamefont {M.}~\bibnamefont
  {Bocquet}}, \bibinfo {author} {\bibfnamefont {J.}~\bibnamefont {Brajard}},
  \bibinfo {author} {\bibfnamefont {A.}~\bibnamefont {Carrassi}}, \ and\
  \bibinfo {author} {\bibfnamefont {L.}~\bibnamefont {Bertino}},\ }\bibfield
  {title} {\enquote {\bibinfo {title} {Bayesian inference of chaotic dynamics
  by merging data assimilation, machine learning and
  expectation-maximization},}\ }\href {\doibase 10.3934/fods.2020004}
  {\bibfield  {journal} {\bibinfo  {journal} {Foundations of Data Science}\
  }\textbf {\bibinfo {volume} {2}},\ \bibinfo {pages} {55} (\bibinfo {year}
  {2020})}\BibitemShut {NoStop}%
\bibitem [{\citenamefont {Rawlins}\ \emph {et~al.}(2007)\citenamefont
  {Rawlins}, \citenamefont {Ballard}, \citenamefont {Bovis}, \citenamefont
  {Clayton}, \citenamefont {Li}, \citenamefont {Inverarity}, \citenamefont
  {Lorenc},\ and\ \citenamefont {Payne}}]{Rawlins07}%
  \BibitemOpen
  \bibfield  {author} {\bibinfo {author} {\bibfnamefont {F.}~\bibnamefont
  {Rawlins}}, \bibinfo {author} {\bibfnamefont {S.~P.}\ \bibnamefont
  {Ballard}}, \bibinfo {author} {\bibfnamefont {K.~J.}\ \bibnamefont {Bovis}},
  \bibinfo {author} {\bibfnamefont {A.~M.}\ \bibnamefont {Clayton}}, \bibinfo
  {author} {\bibfnamefont {D.}~\bibnamefont {Li}}, \bibinfo {author}
  {\bibfnamefont {G.~W.}\ \bibnamefont {Inverarity}}, \bibinfo {author}
  {\bibfnamefont {A.~C.}\ \bibnamefont {Lorenc}}, \ and\ \bibinfo {author}
  {\bibfnamefont {T.~J.}\ \bibnamefont {Payne}},\ }\bibfield  {title} {\enquote
  {\bibinfo {title} {The met office global four-dimensional variational data
  assimilation scheme},}\ }\href {\doibase 10.1002/qj.32} {\bibfield  {journal}
  {\bibinfo  {journal} {Quarterly Journal of the Royal Meteorological Society}\
  }\textbf {\bibinfo {volume} {133}},\ \bibinfo {pages} {347--362} (\bibinfo
  {year} {2007})}\BibitemShut {NoStop}%
\bibitem [{\citenamefont {Wang}, \citenamefont {Wang},\ and\ \citenamefont
  {Zaki}(2019)}]{wang_discrete_2019}%
  \BibitemOpen
  \bibfield  {author} {\bibinfo {author} {\bibfnamefont {M.}~\bibnamefont
  {Wang}}, \bibinfo {author} {\bibfnamefont {Q.}~\bibnamefont {Wang}}, \ and\
  \bibinfo {author} {\bibfnamefont {T.~A.}\ \bibnamefont {Zaki}},\ }\bibfield
  {title} {\enquote {\bibinfo {title} {Discrete adjoint of fractional-step
  incompressible {Navier}-{Stokes} solver in curvilinear coordinates and
  application to data assimilation},}\ }\href {\doibase
  10.1016/j.jcp.2019.06.065} {\bibfield  {journal} {\bibinfo  {journal}
  {Journal of Computational Physics}\ }\textbf {\bibinfo {volume} {396}},\
  \bibinfo {pages} {427--450} (\bibinfo {year} {2019})}\BibitemShut {NoStop}%
\bibitem [{\citenamefont {Anderson}(2001)}]{Anderson01}%
  \BibitemOpen
  \bibfield  {author} {\bibinfo {author} {\bibfnamefont {J.~L.}\ \bibnamefont
  {Anderson}},\ }\bibfield  {title} {\enquote {\bibinfo {title} {An {Ensemble}
  {Adjustment} {Kalman} {Filter} for {Data} {Assimilation}},}\ }\href {\doibase
  10.1175/1520-0493(2001)129<2884:AEAKFF>2.0.CO;2} {\bibfield  {journal}
  {\bibinfo  {journal} {Monthly Weather Review}\ }\textbf {\bibinfo {volume}
  {129}},\ \bibinfo {pages} {2884--2903} (\bibinfo {year} {2001})}\BibitemShut
  {NoStop}%
\bibitem [{\citenamefont {Ruiz}, \citenamefont {Pulido},\ and\ \citenamefont
  {Miyoshi}(2013)}]{Ruiz13}%
  \BibitemOpen
  \bibfield  {author} {\bibinfo {author} {\bibfnamefont {J.~J.}\ \bibnamefont
  {Ruiz}}, \bibinfo {author} {\bibfnamefont {M.}~\bibnamefont {Pulido}}, \ and\
  \bibinfo {author} {\bibfnamefont {T.}~\bibnamefont {Miyoshi}},\ }\bibfield
  {title} {\enquote {\bibinfo {title} {Estimating {Model} {Parameters} with
  {Ensemble}-{Based} {Data} {Assimilation}: {A} {Review}},}\ }\href {\doibase
  10.2151/jmsj.2013-201} {\bibfield  {journal} {\bibinfo  {journal} {Journal of
  the Meteorological Society of Japan. Ser. II}\ }\textbf {\bibinfo {volume}
  {91}},\ \bibinfo {pages} {79--99} (\bibinfo {year} {2013})}\BibitemShut
  {NoStop}%
\bibitem [{\citenamefont {Mons}, \citenamefont {Wang},\ and\ \citenamefont
  {Zaki}(2019)}]{Mons19}%
  \BibitemOpen
  \bibfield  {author} {\bibinfo {author} {\bibfnamefont {V.}~\bibnamefont
  {Mons}}, \bibinfo {author} {\bibfnamefont {Q.}~\bibnamefont {Wang}}, \ and\
  \bibinfo {author} {\bibfnamefont {T.~A.}\ \bibnamefont {Zaki}},\ }\bibfield
  {title} {\enquote {\bibinfo {title} {Kriging-enhanced ensemble variational
  data assimilation for scalar-source identification in turbulent
  environments},}\ }\href {\doibase 10.1016/j.jcp.2019.07.054} {\bibfield
  {journal} {\bibinfo  {journal} {Journal of Computational Physics}\ }\textbf
  {\bibinfo {volume} {398}},\ \bibinfo {pages} {108856} (\bibinfo {year}
  {2019})}\BibitemShut {NoStop}%
\bibitem [{\citenamefont {Carrassi}\ \emph {et~al.}(2008)\citenamefont
  {Carrassi}, \citenamefont {Ghil}, \citenamefont {Trevisan},\ and\
  \citenamefont {Uboldi}}]{Carrassi08}%
  \BibitemOpen
  \bibfield  {author} {\bibinfo {author} {\bibfnamefont {A.}~\bibnamefont
  {Carrassi}}, \bibinfo {author} {\bibfnamefont {M.}~\bibnamefont {Ghil}},
  \bibinfo {author} {\bibfnamefont {A.}~\bibnamefont {Trevisan}}, \ and\
  \bibinfo {author} {\bibfnamefont {F.}~\bibnamefont {Uboldi}},\ }\bibfield
  {title} {\enquote {\bibinfo {title} {Data assimilation as a nonlinear
  dynamical systems problem: Stability and convergence of the
  prediction-assimilation system},}\ }\href {\doibase 10.1063/1.2909862}
  {\bibfield  {journal} {\bibinfo  {journal} {Chaos: An Interdisciplinary
  Journal of Nonlinear Science}\ }\textbf {\bibinfo {volume} {18}},\ \bibinfo
  {pages} {023112} (\bibinfo {year} {2008})}\BibitemShut {NoStop}%
\bibitem [{\citenamefont {Lalescu}, \citenamefont {Meneveau},\ and\
  \citenamefont {Eyink}(2013)}]{Lalescu13}%
  \BibitemOpen
  \bibfield  {author} {\bibinfo {author} {\bibfnamefont {C.~C.}\ \bibnamefont
  {Lalescu}}, \bibinfo {author} {\bibfnamefont {C.}~\bibnamefont {Meneveau}}, \
  and\ \bibinfo {author} {\bibfnamefont {G.~L.}\ \bibnamefont {Eyink}},\
  }\bibfield  {title} {\enquote {\bibinfo {title} {Synchronization of chaos in
  fully developed turbulence},}\ }\href {\doibase
  10.1103/PhysRevLett.110.084102} {\bibfield  {journal} {\bibinfo  {journal}
  {Physical Review Letters}\ }\textbf {\bibinfo {volume} {110}},\ \bibinfo
  {pages} {084102} (\bibinfo {year} {2013})}\BibitemShut {NoStop}%
\bibitem [{\citenamefont {Hoke}\ and\ \citenamefont
  {Anthes}(1976)}]{hoke1976init}%
  \BibitemOpen
  \bibfield  {author} {\bibinfo {author} {\bibfnamefont {J.~E.}\ \bibnamefont
  {Hoke}}\ and\ \bibinfo {author} {\bibfnamefont {R.~A.}\ \bibnamefont
  {Anthes}},\ }\bibfield  {title} {\enquote {\bibinfo {title} {The
  initialization of numerical models by a dynamic-initialization technique},}\
  }\href@noop {} {\bibfield  {journal} {\bibinfo  {journal} {Monthly Weather
  Review}\ }\textbf {\bibinfo {volume} {104}},\ \bibinfo {pages} {1551--1556}
  (\bibinfo {year} {1976})}\BibitemShut {NoStop}%
\bibitem [{\citenamefont {Lakshmivarahan}\ and\ \citenamefont
  {Lewis}(2013)}]{lakshmivarahan2013nudging}%
  \BibitemOpen
  \bibfield  {author} {\bibinfo {author} {\bibfnamefont {S.}~\bibnamefont
  {Lakshmivarahan}}\ and\ \bibinfo {author} {\bibfnamefont {J.~M.}\
  \bibnamefont {Lewis}},\ }\bibfield  {title} {\enquote {\bibinfo {title}
  {Nudging methods: A critical overview},}\ }in\ \href@noop {} {\emph {\bibinfo
  {booktitle} {Data Assimilation for Atmospheric, Oceanic and Hydrologic
  Applications (Vol. II)}}}\ (\bibinfo  {publisher} {Springer},\ \bibinfo
  {year} {2013})\ pp.\ \bibinfo {pages} {27--57}\BibitemShut {NoStop}%
\bibitem [{\citenamefont {Di~Leoni}, \citenamefont {Mazzino},\ and\
  \citenamefont {Biferale}(2020)}]{clark2020synch}%
  \BibitemOpen
  \bibfield  {author} {\bibinfo {author} {\bibfnamefont {P.~C.}\ \bibnamefont
  {Di~Leoni}}, \bibinfo {author} {\bibfnamefont {A.}~\bibnamefont {Mazzino}}, \
  and\ \bibinfo {author} {\bibfnamefont {L.}~\bibnamefont {Biferale}},\
  }\bibfield  {title} {\enquote {\bibinfo {title} {Synchronization to big data:
  Nudging the navier-stokes equations for data assimilation of turbulent
  flows},}\ }\href@noop {} {\bibfield  {journal} {\bibinfo  {journal} {Physical
  Review X}\ }\textbf {\bibinfo {volume} {10}},\ \bibinfo {pages} {011023}
  (\bibinfo {year} {2020})}\BibitemShut {NoStop}%
\bibitem [{\citenamefont {Kalnay}(2003)}]{Kalnay}%
  \BibitemOpen
  \bibfield  {author} {\bibinfo {author} {\bibfnamefont {E.}~\bibnamefont
  {Kalnay}},\ }\href@noop {} {\emph {\bibinfo {title} {Atmospheric Modeling,
  Data Assimilation and Predictability}}}\ (\bibinfo  {publisher} {Cambridge
  University Press},\ \bibinfo {year} {2003})\BibitemShut {NoStop}%
\bibitem [{\citenamefont {Bauer}, \citenamefont {Thorpe},\ and\ \citenamefont
  {Brunet}(2015)}]{Bauer15}%
  \BibitemOpen
  \bibfield  {author} {\bibinfo {author} {\bibfnamefont {P.}~\bibnamefont
  {Bauer}}, \bibinfo {author} {\bibfnamefont {A.}~\bibnamefont {Thorpe}}, \
  and\ \bibinfo {author} {\bibfnamefont {G.}~\bibnamefont {Brunet}},\
  }\bibfield  {title} {\enquote {\bibinfo {title} {The quiet revolution of
  numerical weather prediction},}\ }\href {\doibase 10.1038/nature14956}
  {\bibfield  {journal} {\bibinfo  {journal} {Nature}\ }\textbf {\bibinfo
  {volume} {525}},\ \bibinfo {pages} {47--55} (\bibinfo {year}
  {2015})}\BibitemShut {NoStop}%
\bibitem [{\citenamefont {von Storch}, \citenamefont {Langenberg},\ and\
  \citenamefont {Feser}(2000)}]{Vonstorch00}%
  \BibitemOpen
  \bibfield  {author} {\bibinfo {author} {\bibfnamefont {H.}~\bibnamefont {von
  Storch}}, \bibinfo {author} {\bibfnamefont {H.}~\bibnamefont {Langenberg}}, \
  and\ \bibinfo {author} {\bibfnamefont {F.}~\bibnamefont {Feser}},\ }\bibfield
   {title} {\enquote {\bibinfo {title} {A {Spectral} {Nudging} {Technique} for
  {Dynamical} {Downscaling} {Purposes}},}\ }\href {\doibase
  10.1175/1520-0493(2000)128<3664:ASNTFD>2.0.CO;2} {\bibfield  {journal}
  {\bibinfo  {journal} {Monthly Weather Review}\ }\textbf {\bibinfo {volume}
  {128}},\ \bibinfo {pages} {3664--3673} (\bibinfo {year} {2000})}\BibitemShut
  {NoStop}%
\bibitem [{\citenamefont {Waldron}, \citenamefont {Paegle},\ and\ \citenamefont
  {Horel}(1996)}]{Waldron96}%
  \BibitemOpen
  \bibfield  {author} {\bibinfo {author} {\bibfnamefont {K.~M.}\ \bibnamefont
  {Waldron}}, \bibinfo {author} {\bibfnamefont {J.}~\bibnamefont {Paegle}}, \
  and\ \bibinfo {author} {\bibfnamefont {J.~D.}\ \bibnamefont {Horel}},\
  }\bibfield  {title} {\enquote {\bibinfo {title} {Sensitivity of a
  {Spectrally} {Filtered} and {Nudged} {Limited}-{Area} {Model} to {Outer}
  {Model} {Options}},}\ }\href {\doibase
  10.1175/1520-0493(1996)124<0529:SOASFA>2.0.CO;2} {\bibfield  {journal}
  {\bibinfo  {journal} {Monthly Weather Review}\ }\textbf {\bibinfo {volume}
  {124}},\ \bibinfo {pages} {529--547} (\bibinfo {year} {1996})}\BibitemShut
  {NoStop}%
\bibitem [{\citenamefont {Miguez-Macho}, \citenamefont {Stenchikov},\ and\
  \citenamefont {Robock}(2004)}]{Miguez-macho04}%
  \BibitemOpen
  \bibfield  {author} {\bibinfo {author} {\bibfnamefont {G.}~\bibnamefont
  {Miguez-Macho}}, \bibinfo {author} {\bibfnamefont {G.~L.}\ \bibnamefont
  {Stenchikov}}, \ and\ \bibinfo {author} {\bibfnamefont {A.}~\bibnamefont
  {Robock}},\ }\bibfield  {title} {\enquote {\bibinfo {title} {Spectral nudging
  to eliminate the effects of domain position and geometry in regional climate
  model simulations},}\ }\href {\doibase 10.1029/2003JD004495} {\bibfield
  {journal} {\bibinfo  {journal} {Journal of Geophysical Research:
  Atmospheres}\ }\textbf {\bibinfo {volume} {109}},\ \bibinfo {pages} {D13104}
  (\bibinfo {year} {2004})}\BibitemShut {NoStop}%
\bibitem [{\citenamefont {Biswas}\ \emph {et~al.}(2017)\citenamefont {Biswas},
  \citenamefont {Foias}, \citenamefont {Mondaini},\ and\ \citenamefont
  {Titi}}]{Biswas17}%
  \BibitemOpen
  \bibfield  {author} {\bibinfo {author} {\bibfnamefont {A.}~\bibnamefont
  {Biswas}}, \bibinfo {author} {\bibfnamefont {C.}~\bibnamefont {Foias}},
  \bibinfo {author} {\bibfnamefont {C.~F.}\ \bibnamefont {Mondaini}}, \ and\
  \bibinfo {author} {\bibfnamefont {E.~S.}\ \bibnamefont {Titi}},\ }\bibfield
  {title} {\enquote {\bibinfo {title} {Downscaling data assimilation algorithm
  with applications to statistical solutions of the navier-stokes equations},}\
  }\href {http://arxiv.org/abs/1711.04067} {\bibfield  {journal} {\bibinfo
  {journal} {{arXiv}:1711.04067 [math]}\ } (\bibinfo {year}
  {2017})}\BibitemShut {NoStop}%
\bibitem [{\citenamefont {Foias}, \citenamefont {Mondaini},\ and\ \citenamefont
  {Titi}(2016)}]{Foias16}%
  \BibitemOpen
  \bibfield  {author} {\bibinfo {author} {\bibfnamefont {C.}~\bibnamefont
  {Foias}}, \bibinfo {author} {\bibfnamefont {C.}~\bibnamefont {Mondaini}}, \
  and\ \bibinfo {author} {\bibfnamefont {E.}~\bibnamefont {Titi}},\ }\bibfield
  {title} {\enquote {\bibinfo {title} {A discrete data assimilation scheme for
  the solutions of the two-dimensional navier--stokes equations and their
  statistics},}\ }\href {\doibase 10.1137/16M1076526} {\bibfield  {journal}
  {\bibinfo  {journal} {{SIAM} Journal on Applied Dynamical Systems}\ }\textbf
  {\bibinfo {volume} {15}},\ \bibinfo {pages} {2109--2142} (\bibinfo {year}
  {2016})}\BibitemShut {NoStop}%
\bibitem [{\citenamefont {Albanez}\ \emph {et~al.}(2016)\citenamefont
  {Albanez}, \citenamefont {Lopes}, \citenamefont {J},\ and\ \citenamefont
  {Titi}}]{Albanez16}%
  \BibitemOpen
  \bibfield  {author} {\bibinfo {author} {\bibfnamefont {D.~A.~F.}\
  \bibnamefont {Albanez}}, \bibinfo {author} {\bibfnamefont {N.}~\bibnamefont
  {Lopes}}, \bibinfo {author} {\bibfnamefont {H.}~\bibnamefont {J}}, \ and\
  \bibinfo {author} {\bibfnamefont {E.~S.}\ \bibnamefont {Titi}},\ }\bibfield
  {title} {\enquote {\bibinfo {title} {Continuous data assimilation for the
  three-dimensional {Navier}–{Stokes}-$\alpha$ model},}\ }\href {\doibase
  10.3233/ASY-151351} {\bibfield  {journal} {\bibinfo  {journal} {Asymptotic
  Analysis}\ }\textbf {\bibinfo {volume} {97}},\ \bibinfo {pages} {139--164}
  (\bibinfo {year} {2016})}\BibitemShut {NoStop}%
\bibitem [{\citenamefont {Paz\'o}, \citenamefont {Carrassi},\ and\
  \citenamefont {López}(2016)}]{Pazo16}%
  \BibitemOpen
  \bibfield  {author} {\bibinfo {author} {\bibfnamefont {D.}~\bibnamefont
  {Paz\'o}}, \bibinfo {author} {\bibfnamefont {A.}~\bibnamefont {Carrassi}}, \
  and\ \bibinfo {author} {\bibfnamefont {J.~M.}\ \bibnamefont {López}},\
  }\bibfield  {title} {\enquote {\bibinfo {title} {Data assimilation by
  delay-coordinate nudging},}\ }\href {\doibase 10.1002/qj.2732} {\bibfield
  {journal} {\bibinfo  {journal} {Quarterly Journal of the Royal Meteorological
  Society}\ }\textbf {\bibinfo {volume} {142}},\ \bibinfo {pages} {1290--1299}
  (\bibinfo {year} {2016})}\BibitemShut {NoStop}%
\bibitem [{\citenamefont {Farhat}\ \emph {et~al.}(2019)\citenamefont {Farhat},
  \citenamefont {Glatt-Holtz}, \citenamefont {Martinez}, \citenamefont
  {{McQuarrie}},\ and\ \citenamefont {Whitehead}}]{Farhat19}%
  \BibitemOpen
  \bibfield  {author} {\bibinfo {author} {\bibfnamefont {A.}~\bibnamefont
  {Farhat}}, \bibinfo {author} {\bibfnamefont {N.~E.}\ \bibnamefont
  {Glatt-Holtz}}, \bibinfo {author} {\bibfnamefont {V.~R.}\ \bibnamefont
  {Martinez}}, \bibinfo {author} {\bibfnamefont {S.~A.}\ \bibnamefont
  {{McQuarrie}}}, \ and\ \bibinfo {author} {\bibfnamefont {J.~P.}\ \bibnamefont
  {Whitehead}},\ }\bibfield  {title} {\enquote {\bibinfo {title} {Data
  assimilation in large-prandtl rayleigh-b\'enard convection from thermal
  measurements},}\ }\href {http://arxiv.org/abs/1903.01508} {\bibfield
  {journal} {\bibinfo  {journal} {{arXiv}:1903.01508 [physics]}\ } (\bibinfo
  {year} {2019})}\BibitemShut {NoStop}%
\bibitem [{\citenamefont {Di~Leoni}, \citenamefont {Mazzino},\ and\
  \citenamefont {Biferale}(2018)}]{clark2019}%
  \BibitemOpen
  \bibfield  {author} {\bibinfo {author} {\bibfnamefont {P.~C.}\ \bibnamefont
  {Di~Leoni}}, \bibinfo {author} {\bibfnamefont {A.}~\bibnamefont {Mazzino}}, \
  and\ \bibinfo {author} {\bibfnamefont {L.}~\bibnamefont {Biferale}},\
  }\bibfield  {title} {\enquote {\bibinfo {title} {Inferring flow parameters
  and turbulent configuration with physics-informed data assimilation and
  spectral nudging},}\ }\href@noop {} {\bibfield  {journal} {\bibinfo
  {journal} {Physical Review Fluids}\ }\textbf {\bibinfo {volume} {3}},\
  \bibinfo {pages} {104604} (\bibinfo {year} {2018})}\BibitemShut {NoStop}%
\bibitem [{\citenamefont {Karlin}, \citenamefont {Ferrante},\ and\
  \citenamefont {{\"{O}}ttinger}(1999)}]{Karlin1999}%
  \BibitemOpen
  \bibfield  {author} {\bibinfo {author} {\bibfnamefont {I.~V.}\ \bibnamefont
  {Karlin}}, \bibinfo {author} {\bibfnamefont {A.}~\bibnamefont {Ferrante}}, \
  and\ \bibinfo {author} {\bibfnamefont {H.~C.}\ \bibnamefont
  {{\"{O}}ttinger}},\ }\bibfield  {title} {\enquote {\bibinfo {title} {{Perfect
  entropy functions of the {Lattice} {Boltzmann} method}},}\ }\href@noop {}
  {\bibfield  {journal} {\bibinfo  {journal} {Europhysics Letters (EPL)}\
  }\textbf {\bibinfo {volume} {47}},\ \bibinfo {pages} {182--188} (\bibinfo
  {year} {1999})}\BibitemShut {NoStop}%
\bibitem [{\citenamefont {Ansumali}\ and\ \citenamefont
  {Karlin}(2002)}]{ansumali2002single}%
  \BibitemOpen
  \bibfield  {author} {\bibinfo {author} {\bibfnamefont {S.}~\bibnamefont
  {Ansumali}}\ and\ \bibinfo {author} {\bibfnamefont {I.~V.}\ \bibnamefont
  {Karlin}},\ }\bibfield  {title} {\enquote {\bibinfo {title} {Single
  relaxation time model for entropic lattice boltzmann methods},}\ }\href@noop
  {} {\bibfield  {journal} {\bibinfo  {journal} {Physical Review E}\ }\textbf
  {\bibinfo {volume} {65}},\ \bibinfo {pages} {056312} (\bibinfo {year}
  {2002})}\BibitemShut {NoStop}%
\bibitem [{\citenamefont {Malaspinas}, \citenamefont {Deville},\ and\
  \citenamefont {Chopard}(2008)}]{Malaspinas2008}%
  \BibitemOpen
  \bibfield  {author} {\bibinfo {author} {\bibfnamefont {O.}~\bibnamefont
  {Malaspinas}}, \bibinfo {author} {\bibfnamefont {M.}~\bibnamefont {Deville}},
  \ and\ \bibinfo {author} {\bibfnamefont {B.}~\bibnamefont {Chopard}},\
  }\bibfield  {title} {\enquote {\bibinfo {title} {{Towards a physical
  interpretation of the entropic {Lattice} {Boltzmann} method}},}\ }\href
  {\doibase 10.1103/PhysRevE.78.066705} {\bibfield  {journal} {\bibinfo
  {journal} {Physical Review E}\ }\textbf {\bibinfo {volume} {78}},\ \bibinfo
  {pages} {066705} (\bibinfo {year} {2008})}\BibitemShut {NoStop}%
\bibitem [{\citenamefont {Waleffe}(1992)}]{waleffe1992}%
  \BibitemOpen
  \bibfield  {author} {\bibinfo {author} {\bibfnamefont {F.}~\bibnamefont
  {Waleffe}},\ }\bibfield  {title} {\enquote {\bibinfo {title} {The nature of
  triad interactions in homogeneous turbulence},}\ }\href@noop {} {\bibfield
  {journal} {\bibinfo  {journal} {Physics of Fluids A: Fluid Dynamics}\
  }\textbf {\bibinfo {volume} {4}},\ \bibinfo {pages} {350--363} (\bibinfo
  {year} {1992})}\BibitemShut {NoStop}%
\bibitem [{\citenamefont {Fang}\ \emph {et~al.}(2012)\citenamefont {Fang},
  \citenamefont {Bos}, \citenamefont {Shao},\ and\ \citenamefont
  {Bertoglio}}]{fang2012time}%
  \BibitemOpen
  \bibfield  {author} {\bibinfo {author} {\bibfnamefont {L.}~\bibnamefont
  {Fang}}, \bibinfo {author} {\bibfnamefont {W.~J.}\ \bibnamefont {Bos}},
  \bibinfo {author} {\bibfnamefont {L.}~\bibnamefont {Shao}}, \ and\ \bibinfo
  {author} {\bibfnamefont {J.-P.}\ \bibnamefont {Bertoglio}},\ }\bibfield
  {title} {\enquote {\bibinfo {title} {Time reversibility of navier--stokes
  turbulence and its implication for subgrid scale models},}\ }\href@noop {}
  {\bibfield  {journal} {\bibinfo  {journal} {Journal of Turbulence}\ ,\
  \bibinfo {pages} {N3}} (\bibinfo {year} {2012})}\BibitemShut {NoStop}%
\bibitem [{\citenamefont {Biferale}, \citenamefont {Musacchio},\ and\
  \citenamefont {Toschi}(2012)}]{biferale2012inverse}%
  \BibitemOpen
  \bibfield  {author} {\bibinfo {author} {\bibfnamefont {L.}~\bibnamefont
  {Biferale}}, \bibinfo {author} {\bibfnamefont {S.}~\bibnamefont {Musacchio}},
  \ and\ \bibinfo {author} {\bibfnamefont {F.}~\bibnamefont {Toschi}},\
  }\bibfield  {title} {\enquote {\bibinfo {title} {Inverse energy cascade in
  three-dimensional isotropic turbulence},}\ }\href@noop {} {\bibfield
  {journal} {\bibinfo  {journal} {Physical review letters}\ }\textbf {\bibinfo
  {volume} {108}},\ \bibinfo {pages} {164501} (\bibinfo {year}
  {2012})}\BibitemShut {NoStop}%
\bibitem [{\citenamefont {Chen}, \citenamefont {Chen},\ and\ \citenamefont
  {Eyink}(2003)}]{chen2003joint}%
  \BibitemOpen
  \bibfield  {author} {\bibinfo {author} {\bibfnamefont {Q.}~\bibnamefont
  {Chen}}, \bibinfo {author} {\bibfnamefont {S.}~\bibnamefont {Chen}}, \ and\
  \bibinfo {author} {\bibfnamefont {G.~L.}\ \bibnamefont {Eyink}},\ }\bibfield
  {title} {\enquote {\bibinfo {title} {The joint cascade of energy and helicity
  in three-dimensional turbulence},}\ }\href@noop {} {\bibfield  {journal}
  {\bibinfo  {journal} {Physics of Fluids}\ }\textbf {\bibinfo {volume} {15}},\
  \bibinfo {pages} {361--374} (\bibinfo {year} {2003})}\BibitemShut {NoStop}%
\bibitem [{\citenamefont {Evensen}(2006)}]{Evensen}%
  \BibitemOpen
  \bibfield  {author} {\bibinfo {author} {\bibfnamefont {G.}~\bibnamefont
  {Evensen}},\ }\href@noop {} {\emph {\bibinfo {title} {Data Assimilation: The
  Ensemble Kalman Filter}}}\ (\bibinfo  {publisher} {Springer Science \&
  Business Media},\ \bibinfo {year} {2006})\BibitemShut {NoStop}%
\bibitem [{\citenamefont {Houtekamer}\ and\ \citenamefont
  {Zhang}(2016)}]{Houtekamer16}%
  \BibitemOpen
  \bibfield  {author} {\bibinfo {author} {\bibfnamefont {P.~L.}\ \bibnamefont
  {Houtekamer}}\ and\ \bibinfo {author} {\bibfnamefont {F.}~\bibnamefont
  {Zhang}},\ }\bibfield  {title} {\enquote {\bibinfo {title} {Review of the
  ensemble kalman filter for atmospheric data assimilation},}\ }\href {\doibase
  10.1175/MWR-D-15-0440.1} {\bibfield  {journal} {\bibinfo  {journal} {Monthly
  Weather Review}\ }\textbf {\bibinfo {volume} {144}},\ \bibinfo {pages}
  {4489--4532} (\bibinfo {year} {2016})}\BibitemShut {NoStop}%
\bibitem [{\citenamefont {Talagrand}\ and\ \citenamefont
  {Courtier}(1987)}]{Talagrand87}%
  \BibitemOpen
  \bibfield  {author} {\bibinfo {author} {\bibfnamefont {O.}~\bibnamefont
  {Talagrand}}\ and\ \bibinfo {author} {\bibfnamefont {P.}~\bibnamefont
  {Courtier}},\ }\bibfield  {title} {\enquote {\bibinfo {title} {Variational
  assimilation of meteorological observations with the adjoint vorticity
  equation. i: Theory},}\ }\href {\doibase 10.1002/qj.49711347812} {\bibfield
  {journal} {\bibinfo  {journal} {Quarterly Journal of the Royal Meteorological
  Society}\ }\textbf {\bibinfo {volume} {113}},\ \bibinfo {pages} {1311--1328}
  (\bibinfo {year} {1987})}\BibitemShut {NoStop}%
\end{thebibliography}%

\end{document}